# *Macroscopic Electromagnetic Response of Arbitrarily Shaped Spatially Dispersive Bodies formed by Metallic Wires*

*João T. Costa[(1)], Mário G. Silveirinha[(1)]*

[(1)] *University of Coimbra, Electrical Engineering Department-Instituto de Telecomunicações, 3030-290 Coimbra, Portugal,*

*joao.costa@co.it.pt, mario.silveirinha@co.it.pt*

In media with strong spatial dispersion the electric displacement vector and the electric field are typically linked by a partial differential equation in the bulk region. The objective of this work is to highlight that in the vicinity of an interface the relation between the macroscopic fields cannot be univocally determined from the bulk response of the involved materials, but requires instead the knowledge of internal degrees of freedom of the materials. We derive such a relation for the particular case of "wire media", and describe a numerical formalism that enables characterizing the electromagnetic response of arbitrarily shaped spatially dispersive bodies formed by arrays of crossed wires. The possibility of concentrating the electromagnetic field in a narrow spot by tapering a metamaterial waveguide is discussed.

---

[*] To whom correspondence should be addressed.

# I. Introduction

Spatially dispersive materials have the peculiar property that the macroscopic polarization vector depends not only on the macroscopic electric field, but also on its spatial derivatives [1]. As is well known, this implies that the electric displacement vector **D** is related to the electric field **E** through a constitutive relation of the form $\mathbf{D} = \bar{\bar{\varepsilon}}(\omega, -i\nabla) \cdot \mathbf{E}$, which for the case of fields with a plane wave type spatial dependence of the form $e^{i\mathbf{k}\cdot\mathbf{r}}$ reduces simply to $\mathbf{D} = \bar{\bar{\varepsilon}}(\omega, \mathbf{k}) \cdot \mathbf{E}$. The dielectric function $\bar{\bar{\varepsilon}}(\omega, \mathbf{k})$ fully characterizes the electromagnetic response of the material in the bulk region for any macroscopic excitation. However, any realistic physical system is necessarily of finite extent, and some of the most interesting electromagnetic phenomena – such as the refraction and reflection of light, field localization and waveguiding – have their origin in interface effects. In general, the constitutive relation $\mathbf{D} = \bar{\bar{\varepsilon}}(\omega, -i\nabla) \cdot \mathbf{E}$, does not hold exactly at the boundary, and this can create ambiguities in the solution of electromagnetic problems involving bodies formed by spatially dispersive materials.

To illustrate this, let us consider the simple case where both the electric field and electric displacement field are oriented along the *z*-direction, and are linked in the bulk region as follows:

$$D_z = \varepsilon(\omega, -i\nabla) E_z. \tag{1}$$

Furthermore, for the purpose of illustration it is assumed that the material is non-magnetic and that the dielectric function is a rational function of the wave vector, so that

$$\varepsilon(\omega, \mathbf{k}) = \varepsilon_h + \frac{b_0}{a_0 - a_2 k_x^2 + ...}, \tag{2}$$

where $\varepsilon_h$, $b_0$, $a_0$, $a_2$,..., are independent of the wave vector, but in general may depend on frequency. It is supposed that the material has a center of symmetry at the microscopic level so that the dielectric function is an even function of $\mathbf{k}$. Moreover, it is assumed without loss of generality that $\varepsilon(\omega,\mathbf{k})$ depends exclusively on $k_x \leftrightarrow -i\dfrac{\partial}{\partial x}$. Clearly, in case the only nonzero coefficients are $a_0$, $a_2$, the $D_z$ and $E_z$ fields satisfy the following partial differential equation in the bulk region:

$$a_0 P_{c,z} + \partial_x^2 \left(a_2 P_{c,z}\right) = b_0 E_z. \tag{3}$$

where we defined $P_{c,z} = D_z - \varepsilon_h E_z$ which may be regarded as the polarization of the medium with respect to a background with permittivity $\varepsilon_h$. For a material with a local response the coefficient $a_2$ vanishes.

Let us now suppose that the plane $x = 0$ corresponds to an interface between two different materials, so that one of the materials occupies the semispace $x > 0$, whereas the second material occupies the region $x < 0$, and that the constitutive relation in both bulk materials is of the generic form of Eq. (2). Evidently, the coefficients $a_0$, $a_2$, and $b_0$ in general differ in the two materials. Therefore, it is tempting to consider that the $P_{c,z}$ and $E_z$ fields are related in all space by:

$$a_0(x) P_{c,z} + \partial_x^2 \left(a_2(x) P_{c,z}\right) = b_0(x) E_z. \tag{4}$$

This equation together with the standard macroscopic Maxwell's Equations, $\nabla \times \mathbf{E} = i\omega\mu_0 \mathbf{H}$ and $\nabla \times \mathbf{H} = \mathbf{j}_{ext} - i\omega \mathbf{D}$ and the Sommerfeld radiation conditions completely determine, for a given excitation $\mathbf{j}_{ext}$, the electromagnetic fields $(\mathbf{E},\mathbf{H})$ in all space. The outlined ideas and other variants are the basis of several studies which aim at characterizing the electromagnetic response of either nanoparticles or macroscopic bodies made of either natural media or metamaterials with spatial dispersion [2-10]. The main objective of the present study is to demonstrate with specific examples that even though in some scenarios this direct approach captures correctly the physical response of a system, in other cases it may produce inaccurate results.

Indeed, even if the bulk constitutive relation (3) holds exactly up to the boundary, in general the form of Eq. (4) remains unjustified at $x = 0$, i.e. at the boundary. The reason is that there are many inequivalent ways of relating $P_{c,z}$ and $E_z$ through a differential equation, but which reduce to Eq. (3) in the bulk regions. In fact, since for an abrupt interface the coefficients $a_0$, $a_2$, and $b_0$ are discontinuous at $x = 0$, *a priori* nothing forbids that $P_{c,z}$ and $E_z$ are linked by, for example,

$$a_0(x) P_{c,z} + \partial_x \left[ a_2(x) \partial_x P_{c,z} \right] = b_0(x) E_z, \qquad (5)$$

rather than by Eq. (4). Notice that the above equation is equivalent to Eq. (4) in the bulk regions (i.e. for $x \neq 0$ where $a_2(x) = const.$) but not at $x = 0$. Indeed, the form of Eq. (4) suggests that $\partial_x (a_2 P_{c,z})$ is continuous at the interface, whereas differently Eq. (5) implies that $a_2 \partial_x P_{c,z}$ is continuous at the interface. Hence, the two formulations imply different boundary conditions at the interfaces, even though they are equivalent in the bulk regions. In this discussion, it is implicit that the pertinent solution ($P_{c,z}$) is defined in space of generalized functions, and that the equations hold in the distributional sense. It is interesting to mention that generally, when electromagnetic waves illuminate spatially dispersive bodies, "additional waves" can be excited and hence the classical boundary conditions that impose the continuity of the tangent fields at the interfaces are insufficient to solve a scattering problem based on mode matching. The usual way to fix this problem is to impose additional boundary conditions (ABCs) [1, 11-13]. However, it is important to mention that in a framework where the bulk constitutive relations are extended across the interface (e.g. Eq. (4) or Eq. (5)), the scattering problem is complete and logically consistent on its own, and hence it does not require further boundary conditions to be explicitly imposed. However, as should be clear from the above argument, the structure of adopted constitutive relation (e.g. Eq. (4) or Eq. (5)) at the boundary may indirectly enforce an ABC at an interface where the coefficients of the equation are discontinuous. Therefore the knowledge of the correct form of the constitutive relation across the interface is intimately related to the knowledge of ABCs, and these are complementary aspects of the same problem.

The previous discussion illustrates that there are distinct ways of linking $P_{c,z}$ and $E_z$ close to the boundary, but which are consistent with the constitutive relations in the bulk regions. In fact, there are infinitely many possibilities of linking $P_{c,z}$ and $E_z$ at the boundary, and some of them cannot even be formulated in terms of the coefficients $a_0$, $a_2$, and $b_0$ of the effective medium model! For example, if one replaces the term $\partial_x(a_2 P_{c,z})$ in Eq. (4) by the term $A(x)\partial_x\left[A^{-1}(x)\partial_x(a_2 P_{c,z})\right]$ where $A(x)$ is an *arbitrary* piecewise constant function of $x$ discontinuous at $x=0$, one obtains other inequivalent ways of linking $P_{c,z}$ and $E_z$ in all space, involving an extra parameter ($A(x)$) which is unrelated to the bulk material dielectric function.

In this work, our aims are *(i)* to highlight that the correct manner of extending the bulk constitutive relations across the interface requires the knowledge of internal (microscopic) degrees of freedom of the involved materials at the boundary, and *(ii)* to discuss how the Maxwell's equations can be solved using numerical methods in the presence of arbitrarily shaped bodies with a spatially dispersive response. To this end, we investigate the electromagnetic response of arbitrarily shaped bodies of "wire media", which are metamaterials known to have a strongly spatially dispersive response [14-17] and interesting applications in the emerging fields of nanophotonics and plasmonics [18-26].

The uniaxial wire medium [14] is the most well-known metamaterial with a nonlocal response, but such a property is also inherent to other wire media topologies, such as arrays of long helices and arrays of both connected and nonconnected wires [13, 17]. In this work, we choose the double wire medium – a double array of nonconnected metallic wires – for illustration purposes, but the theory can be trivially extended to other wire medium topologies. Based on an effective medium framework wherein the metamaterial response is expressed in terms of additional variables with known physical meaning [27], we prove that the correct manner of linking the **D** and **E** fields across the boundary does not reduce to a simple Fourier inversion of the bulk constitutive relations as in Eq. (4). We use our theory to develop a spatially dispersive finite-difference frequency-domain (FDFD-SD) numerical method that enables solving the Maxwell-Equations in scenarios wherein electromagnetic waves can interact with arbitrarily shaped bodies formed by wire media. We demonstrate with numerical simulations that if the host medium of the metallic wires is a dielectric, or even more drastic, if the nanowires are in contact with a metallic surface, a numerical solution based on Eq. (4) may fail at the interfaces. We apply the FDFD-SD formalism to investigate applications of the "double wire medium" in superlensing [18] and in ultraconfined waveguiding. In this work, a time variation of the form $e^{-i\omega t}$ is assumed.

## II. Model based on the Bulk Electromagnetic Response

As mentioned in the Introduction, without loss of generality this work is focused in the electromagnetic response of the double wire medium. This material is formed by two arrays of metallic wires with radius $r_w$, such that each array of parallel wires is arranged in a square lattice with lattice constant $a$ and tilted by ±45° with respect to the interfaces.

One set of wires is oriented along the direction $\hat{\mathbf{u}}_1 = (1,0,1)/\sqrt{2}$ while the complementary set of wires is oriented along the direction $\hat{\mathbf{u}}_2 = (-1,0,1)/\sqrt{2}$. Both sets of wires lie in planes parallel to the *xoz* plane and the distance between adjacent perpendicular wires is $a/2$ [Figs. 1a and 1b].

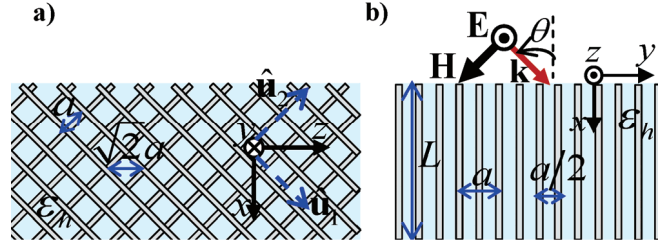

Fig. 1. (Color online) (a) and (b) show cuts of a "double wire medium" along the *xoy* and *xoz* planes, respectively. The slab has thickness $L$.

The wires stand in a host material with relative permittivity $\varepsilon_h$. The effective response of the "double wire medium" is characterized by a dielectric function $\overline{\overline{\varepsilon}}(\omega, \mathbf{k})$ such that [12, 16, 28]:

$$\frac{\overline{\overline{\varepsilon}}}{\varepsilon_0} = \varepsilon_h \hat{\mathbf{u}}_y \hat{\mathbf{u}}_y + \varepsilon_{11} \hat{\mathbf{u}}_1 \hat{\mathbf{u}}_1 + \varepsilon_{22} \hat{\mathbf{u}}_2 \hat{\mathbf{u}}_2$$

$$\varepsilon_{ii}(\omega, k_i) = \varepsilon_h \left( 1 + \frac{1}{\dfrac{1}{(\varepsilon_m/\varepsilon_h - 1)f_V} - \dfrac{\varepsilon_h(\omega/c)^2 - k_i^2}{\beta_p^2}} \right), \tag{6}$$

where $\varepsilon_0 \varepsilon_m$ is the permittivity of the metal, $f_V = \pi(r_w/a)^2$ is the volume fraction of each set of wires, $\beta_p = \{2\pi/[\ln(a/2\pi r_w) + 0.5275]\}^{1/2}/a$ is the plasma wave number and $c$ is the speed of light in vacuum. For simplicity, in this work we restrict our attention to the case of propagation along the *xoy* plane with $k_z = 0$ (or equivalently $\partial_z = 0$), and assume that the only nontrivial electromagnetic field components are $E_z$, $D_z$, $H_x$ and $H_y$. In this scenario, the dielectric function reduces to a scalar in the *xoy* plane, $\varepsilon(\omega, k_x) = \varepsilon_{ii}(\omega, k_i)$, $i = 1, 2$, because for $k_z = 0$ we have $k_1 = k_x/\sqrt{2} = -k_2$. Therefore, in this situation Eq. (6) becomes:

$$\varepsilon(\omega, k_x) = \varepsilon_h \left( 1 + \frac{1}{\frac{1}{(\varepsilon_m/\varepsilon_h - 1)f_V} - \frac{\varepsilon_h(\omega/c)^2 - k_x^2/2}{\beta_p^2}} \right), \qquad (7)$$

which is clearly of the same form as in Eq. (2), i.e., it is a rational function of the wave vector.

Because the only non-zero field components in the problems that we are interested in are $E_z$, $D_z$, $H_x$ and $H_y$, it is easily found that the Maxwell's equations, $\nabla \times \mathbf{E} = i\omega\mu_0 \mathbf{H}$ and $\nabla \times \mathbf{H} = \mathbf{j}_{ext} - i\omega \mathbf{D}$, reduce to the scalar equation:

$$\frac{\partial^2}{\partial x^2} E_z + \frac{\partial^2}{\partial y^2} E_z + \left(\frac{\omega}{c}\right)^2 \frac{D_z}{\varepsilon_0} = -i\omega\mu_0 j_{s,z}, \qquad (8)$$

where $\mathbf{j}_{ext} = j_{s,z}\hat{\mathbf{z}}$ represents an external current density, i.e. an external excitation. Therefore, provided one is able to link $E_z$ and $D_z$ in all space the Maxwell's Equations can be solved univocally. Next, we discuss how this can be done based on Eq. (7).

## *A. Constitutive Relations in the Bulk Region*

Similar to what was outlined in the Introduction, substituting Eq. (7) into Eq. (1) and calculating the inverse Fourier transform ($ik_x \leftrightarrow \partial_x$) of the resulting expression, it is possible to obtain a spatial relation between the electric field $E_z$ and the electric displacement $D_z$ that makes manifest the spatially dispersive nature of the response of the metamaterial:

$$\left[\frac{1}{2}\frac{\partial^2}{\partial x^2} + \varepsilon_h\left(\frac{\omega}{c}\right)^2 + \beta_c^2\right]\frac{P_{c,z}}{\varepsilon_0} + \varepsilon_h\beta_p^2 E_z = 0, \qquad (9)$$

where $\beta_c^2 = -\dfrac{\beta_p^2}{(\varepsilon_m/\varepsilon_h - 1)f_V}$ and $P_{c,z} = D_z - \varepsilon_0 \varepsilon_h E_z$. This is analogous to Eq. (3) for the particular case of the double wire medium. It should be noted that $P_{c,z}$ is the contribution to the polarization vector due to the conduction currents in the nanowires. Thus, Eq. (9) effectively determines the response of the conduction polarization current to the "applied" macroscopic electric field.

It is stressed that *a priori* Eq. (9) is only valid in the bulk region of the metamaterial. However, it can be trivially extended to scenarios wherein a metamaterial body is surrounded by a standard dielectric (let us say air). Indeed, if one regards $\varepsilon_h \equiv \varepsilon_h(x,y)$ as a position dependent function that represents the relative permittivity of the background dielectric regions, and similarly $\beta_p \equiv \beta_p(x,y)$ and $\beta_c \equiv \beta_c(x,y)$ as functions that vanish outside the metamaterial, it is clear that in a standard dielectric Eq. (9) reduces to:

$$\left[ \varepsilon_h \left(\frac{\omega}{c}\right)^2 + \frac{1}{2}\frac{\partial^2}{\partial x^2} \right] P_{c,z} = 0. \tag{10}$$

This relation is correct because in a standard dielectric $D_z = \varepsilon_0 \varepsilon_h E_z$, or equivalently $P_{c,z} = 0$. Thus, if one lets $\varepsilon_h \equiv \varepsilon_h(x,y)$, $\beta_p \equiv \beta_p(x,y)$ and $\beta_c \equiv \beta_c(x,y)$ be space dependent, Eq. (9) yields the correct constitutive relations both in the bulk metamaterial and in the bulk dielectric region (i.e in the region that surrounds the metamaterial body). If in addition one assumes that Eq. (9) also holds across the boundary – which as discussed in Sect. I in general may be a "leap of faith" – then it is possible to calculate the electromagnetic fields in all space by combining and solving Eq. (8) and Eq. (9). In the next sub-section, we briefly describe how this can be done numerically using the FDFD method.

## B. FDFD Discretization

The unknown fields [solution of Eqs. (8) and (9)] can be obtained using the well-known FDFD method based on the Yee's mesh [29]. The discretization of the second order derivatives in these equations is done based on the formulas proposed in [30]:

$$\frac{\partial^2}{\partial x^2} F(i,j) = \frac{F(i+1,j) - 2F(i,j) + F(i-1,j)}{\Delta x^2} \tag{11a}$$

$$\frac{\partial^2}{\partial y^2} F(i,j) = \frac{F(i,j+1) - 2F(i,j) + F(i,j-1)}{\Delta y^2}, \tag{11b}$$

where $F = P_{c,z} = D_z - \varepsilon_0 \varepsilon_h E_z$, $\Delta x$ and $\Delta y$ is the grid spacing along the $x$- and $y$-directions, respectively, and the discrete indices $(i,j)$ stand for a given $i$-th and $j$-th node of the grid mesh along the $x$- and $y$-directions, respectively. As discussed previously, $\varepsilon_h \equiv \varepsilon_h(x,y)$ is a position dependent function that is equal to the host permittivity in the metamaterial and to the permittivity constant in the dielectric material. On the other hand, $\beta_p \equiv \beta_p(x,y)$ and $\beta_c \equiv \beta_c(x,y)$ are set equal to zero outside the metamaterial.

We considered two FDFD solutions for the described problem. *(i)* In the first approach Eq. (9) is used in all the regions of space to link $E_z$ and $D_z$. We will refer to this solution as the direct inverse transform (DIT1) solution. *(ii)* In the second approach we use Eq. (9) to link $E_z$ and $D_z$ inside the metamaterial as well as for all the nodes that are over the boundary. For nodes that are completely outside the metamaterial (and such that all the neighboring nodes are also outside the metamaterial) we use simply $D_z = \varepsilon_0 \varepsilon_h E_z$ rather than Eq. (9). We will refer to this implementation as DIT2. The perfectly matched layer (PML) described in [31] is used to truncate the computation domain in both implementations.

## III. Model based on Internal Degrees of Freedom of the Medium

Recent works [13, 27, 32] have shown that the spatial dispersion inherent to wire media may be described by a quasi-static homogenization model that applies in a wide range of scenarios, including the case where the wires are periodically loaded with conducting metallic bodies. In this homogenization framework a current $I$ and an additional potential $\varphi$, are associated with each set of wires. The current $I$ may be identified with the current that flows along the metallic wires, whereas the additional potential is the average quasi-static potential drop from a given wire to the boundary of the respective unit cell (both the current and the additional potential are interpolated in a suitable manner, so that they become continuous functions of the spatial coordinates) [27]. In particular, as detailed in Appendix A, for the case of the double wire medium the electrodynamics of the metamaterial is described by a 10-component state vector $\mathbf{F} = (\mathbf{E}, \mathbf{H}, \varphi_1, I_1, \varphi_2, I_2)$ that satisfies a differential system of the form:

$$\hat{\mathbf{L}} \cdot \mathbf{F} = +i\omega \hat{\mathbf{M}} \cdot \mathbf{F} + \mathbf{J}_{ext}, \qquad (12)$$

where $\hat{\mathbf{L}} = \hat{\mathbf{L}}(\nabla)$ is a linear differential operator, $\hat{\mathbf{M}} = \hat{\mathbf{M}}(\varepsilon_h, \mu_0, L_w, Z_w, C_w)$ is a material matrix that depends on the geometry of the array of metallic wires and on the electromagnetic properties of the involved materials, and $\mathbf{J}_{ext}$ represents a source term. The important point is that this formalism based on the introduction of additional variables provides a framework where the wire medium response is "local" (even though the electrodynamics is nonlocal) in the sense that that the material response can be written in terms of the ten-component vector $\mathbf{F} = (\mathbf{E}, \mathbf{H}, \varphi_1, I_1, \varphi_2, I_2)$ through a linear operator ($\hat{\mathbf{M}}$) independent of the spatial derivatives. Hence, it is reasonable to assume that Eq. (12) holds even across a boundary between two materials with different structural parameters, such that $\hat{\mathbf{M}} = \hat{\mathbf{M}}(x, y, z)$. Such premise will be the basis of the ideas developed in this section, where we obtain a solution for the electromagnetic fields in all space relying on Eq. (12). Notice that a standard dielectric can also be described with this formalism since it can be considered as the limit of a nanowire material with vanishingly thin wires.

It is important to emphasize that the effective medium formalism associated with Eq. (12) is based on the knowledge of the dynamics of the additional variables $I$ and $\varphi$, which have known physical meaning, and thus is based on the knowledge of internal degrees of freedom of the material.

## A. Constitutive relations based on the internal degrees of freedom

In Appendix A, it is shown that in the general case where the structural parameters are arbitrary functions of the coordinates ($\hat{M}=\hat{M}(x,y,z)$), Eq. (12) reduces to:

$$\frac{\varepsilon_h \beta_p^2}{2} \frac{\partial}{\partial x}\left[\frac{1}{\varepsilon_h \beta_p^2} \frac{\partial}{\partial x} \frac{P_{c,z}}{\varepsilon_0}\right] + \left[\varepsilon_h \left(\frac{\omega}{c}\right)^2 + \beta_c^2\right]\frac{P_{c,z}}{\varepsilon_0} + \varepsilon_h \beta_p^2 E_z = 0, \qquad (13a)$$

$$\frac{\partial^2}{\partial x^2} E_z + \frac{\partial^2}{\partial y^2} E_z + \left(\frac{\omega}{c}\right)^2 \frac{D_z}{\varepsilon_0} = -i\omega\mu_0 j_{s,z}. \qquad (13b)$$

Evidently, Eq. (13b) is the same as Eq. (8). On the other hand Eq. (13a) is precisely the same as Eq. (9) in the bulk region, i.e., when $\beta_p$ and $\varepsilon_h$ are constant and independent of the position. However, the two equations are completely different at the interfaces, since the parameters $\beta_p$, $\beta_c$ and $\varepsilon_h$ may vary with space. This happens if for example the permittivity of the host medium or the radii of the wires vary in space.

In the same manner as in Sect. II, here we assume that $\varepsilon_h = \varepsilon_h(x,y)$, $\beta_p = \beta_p(x,y)$ and $\beta_c = \beta_c(x,y)$. In a standard dielectric, we take the limit $\beta_p \to 0$ and put $\beta_c(x,y) = 0$. Note that in this case $\beta_p$ cannot be chosen exactly equal to zero, otherwise Eq. (13a) becomes singular.

## B. FDFD Discretization

The FDFD discretization of the system (13) is analogous to that already described in Sect. II.B. The only relevant difference is that the second order derivatives of Eqs. (13) are of the generic form $\frac{\partial}{\partial x} G(x,y) \frac{\partial}{\partial x} U(x,y)$, where $G(x,y) = 1/\left[\varepsilon_h(x,y)\beta_p^2(x,y)\right]$ and $U(x,y) = P_{c,z}/\varepsilon_0$. The derivative $\frac{\partial}{\partial x} G(x,y) \frac{\partial}{\partial x} U(x,y)$ is discretized in the following manner:

$$\left[\frac{\partial}{\partial x} G \frac{\partial}{\partial x} U\right](i,j) = \frac{A(i,j)U(i+1,j)}{\Delta x^2} - \frac{B(i,j)U(i,j)}{\Delta x^2} + \frac{C(i,j)U(i-1,j)}{\Delta x^2}, \quad (14)$$

where $A(i,j) = \left[G(i,j) + G(i+1,j)\right]/2$, $B(i,j) = G(i,j) + \left[G(i+1,j) + G(i-1,j)\right]/2$, $C(i,j) = \left[G(i,j) + G(i-1,j)\right]/2$. The computation domain is truncated with a PML [31]. In this implementation, we use (13a) in all space (both in the metamaterial and in standard dielectrics or metals). We will refer to this solution based on the internal degrees of freedom of the metamaterial as the "IDF solution". The discretized Eqs (13) are given in Appendix B.

## IV. Numerical Results and Discussion

Next, we compare the results obtained with the formulations of Secs. II and III and confirm that the form of the constitutive relations at the interfaces is of crucial importance.

In the first example, we consider a double wire medium slab formed by PEC wires, i.e, $\varepsilon_m = -\infty$. The metamaterial has thickness $L$ and is surrounded by air (Fig. 1b). The permittivity of the host region in the double wire medium is taken equal to $\varepsilon_h = 10$, and the lattice constant $a$ is such that $a = L/20$ and $r_w = 0.05a$. In Fig. 2 the reflection and transmission coefficients $\rho$ and $\tau$ are depicted as a function of the normalized frequency $\omega L / c$ for a plane wave that illuminates the slab with an angle of incidence $\theta_i = 15°$.

The green triangles and the blue circles represent the results computed with the FDFD-SD methods DIT1 and DIT2, respectively (see Sect. II.B). These two approaches are based on Eq. (9). On the other hand, the orange stars were obtained using the FDFD-SD method IDF (see Sect. III.B) based on the knowledge of the internal structure of the metamaterial [Eq. (13)]. Note that in the implementations DIT1 and IDF the parameter $\beta_p^2$ is taken as vanishingly small outside the metamaterial. Finally, the black solid curves in Fig. 2 were computed using an analytical approach derived in Ref. [12], based on mode matching and additional boundary conditions. It was demonstrated in Ref. [12] (this is further confirmed in Fig. 3) that this analytical method compares very well with full wave simulations that take into account all the minute details of the microstructure of the metamaterial. Therefore, the solid curves can be regarded here as the "exact solution" of the problem.

Figure 2 shows that the DIT1 method can be quite inaccurate, as the green curve for the amplitude of the transmission coefficient $\tau$ (Fig. 2c) largely mismatches the curve obtained with the analytical model (solid black curve). This confirms that a proper discretization of the electromagnetic fields at the interfaces between the spatially dispersive metamaterial and the air region is of crucial importance. On the other hand, the blue curves (DIT2) concur better with the analytical model. The results obtained with the IDF implementation (orange symbols) yield a nearly perfect agreement with the analytical formalism. This supports that to model correctly the electromagnetic response of spatially dispersive bodies it may be necessary to know some of the internal degrees of freedom of the metamaterial, which cannot be accessed simply from the knowledge of the bulk electromagnetic response. In Fig. 3 we compare the results obtained with the IDF implementation and full-wave simulations [33] that take into account the microstructure of the metamaterial. As seen, the agreement is nearly perfect.

The several dips in the reflection characteristic in Fig. 2 are associated with Fabry-Pérot resonances. These resonances are ultra-subwavelength (e.g. the first dip of the reflection coefficient occurs at $\omega L/c \approx 0.13$, which corresponds to the metallic wires with length $L_{wm} = \sqrt{2}L = 0.03\lambda_0$) because the "double wire medium" can be characterized by a very large positive index of refraction with anomalous frequency dispersion in the low frequency limit [21, 26]. It is interesting to mention that the electromagnetic response of "wire media" has typically a dual behavior, so that depending on the excitation the metamaterial may behave as either an effective medium with positive permittivity or as a material with negative permittivity [32].

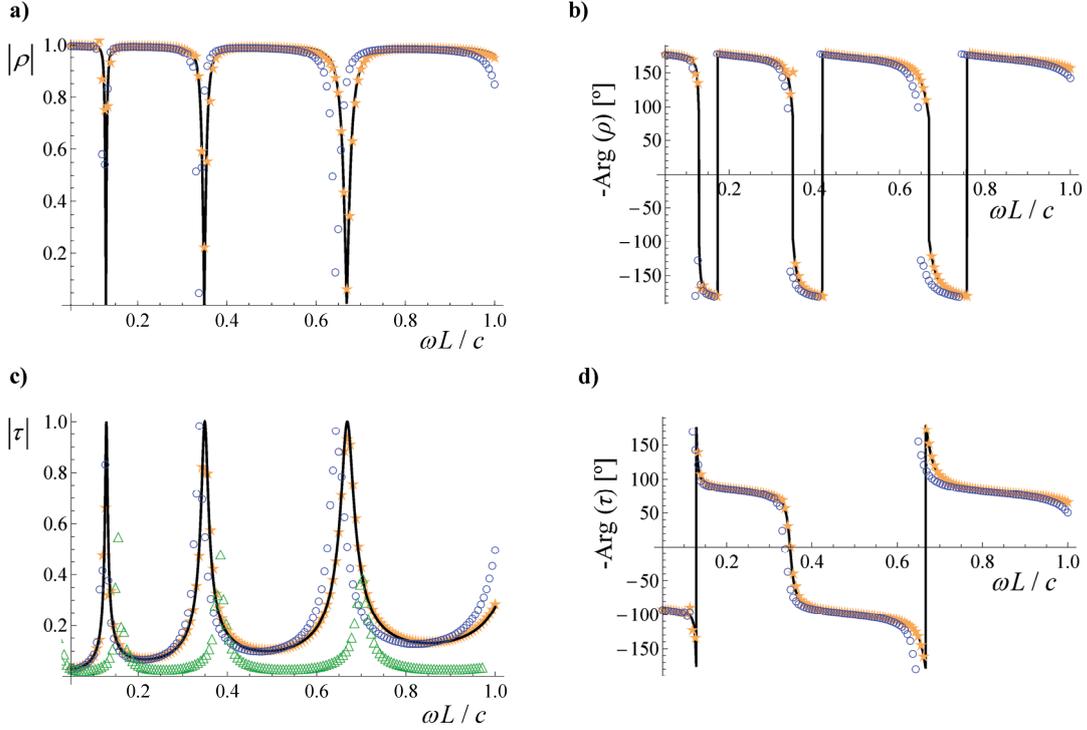

Fig. 2. (Color online) Reflection and transmission coefficients as a function of the normalized frequency for a double wire medium substrate with thickness $L = 20a$ illuminated by a plane wave with angle of incidence $\theta_i = 15º$. Solid (black) curves: mode-matching approach based on additional boundary conditions [12]. Star shaped (orange) symbols: IDF approach (Sect. III.B); Triangle shaped (green) symbols: DIT1 approach (Sect. II.B); Circle shaped (blue) symbols: DIT2 approach (Sect. II.B); (a) and (b): amplitude and phase of the reflection coefficient $\rho$, respectively. (c) and (d): amplitude and phase of the transmission coefficient $\tau$, respectively.

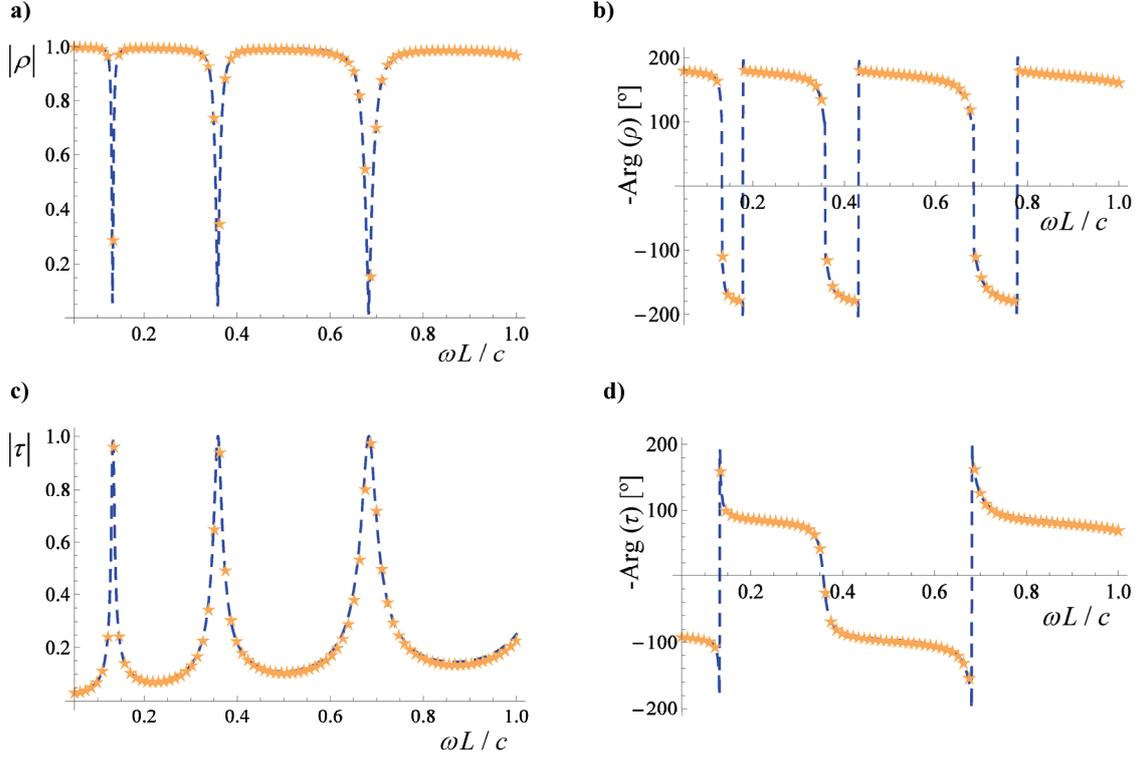

Fig. 3. (Color online) Same example as in Fig. 2. Blue dashed curves: Full wave results obtained with CST Microwave Studio [33]. Star shaped (orange) symbols: IDF approach (Sect. III.B).

To illustrate the application of the method in case of metallic loss, next we suppose that the metal permittivity $\varepsilon_0 \varepsilon_m$ has a Drude-type dispersion so that $\varepsilon_m = 1 - \dfrac{\omega_p^2}{\omega(\omega + i\Gamma)}$, where $\omega_p$ is the plasma frequency and $\Gamma$ is the collision frequency. It is assumed that the plasma frequency is such that $\omega_p a / c = 0.125$ and that the collision frequency is $\Gamma / \omega_p = 0.05$. The remaining structural parameters, as well as the incoming wave, are as in Fig. 2. The reflection and transmission coefficients calculated with the IDF approach and with the analytical (ABC based) approach [12] are plotted in Fig. 4. As seen, the agreement between the FDFD-SD results and the analytical model is excellent, confirming that the constitutive relation (13) is valid across the interfaces between different media even in case of metal loss.

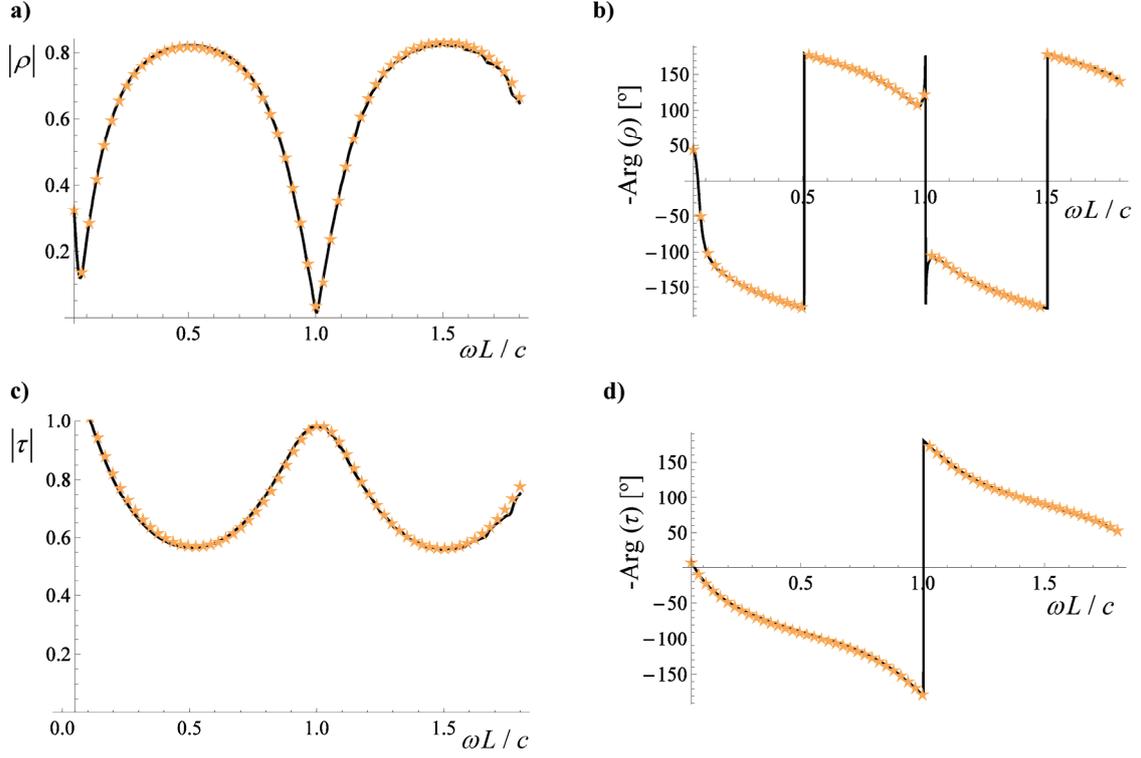

Fig. 4. (Color online) Similar to Fig. 2 but the permittivity of the wires is described by the Drude model $\varepsilon_m = 1 - \omega_p^2 / \omega(\omega + i\Gamma)$. The parameters of the Drude Model are $\omega_p a / c = 0.125$ and $\Gamma / \omega_p = 0.05$. Solid (black) curves: mode-matching approach based on additional boundary conditions [12]. Star shaped (orange) symbols: IDF approach (Sect. III.B);

Next, we consider the case where the metamaterial slab is backed by a metallic region (a very good conductor, which we will refer to as the "ground plane"; the permittivity of the ground plane is taken as $\varepsilon_h \to -\infty$). The incoming wave propagates in air as in the previous examples, and the angle of incidence is taken equal to $\theta_i = 70°$. The thickness of the slab is $L = 20a$, the radius of the wires is $r_w = 0.05a$, and the relative permittivity of the host region is $\varepsilon_h = 30$. The metallic wires are assumed PEC.

We consider the scenario where the metallic wires are in ohmic contact with the ground plane. In the formalism DIT2 (Sect. II.B) there is no way of specifying that the metallic wires are in contact with the "ground plane". On the other hand, in the models DIT1 and IDF this can be taken into account by imagining that the wires are slightly prolonged into the metal, so that they penetrate in a thin transition layer inside the metal. Thus, in the transition layer the parameter $\beta_p$ (which only depends on the radius of the wires) is taken the same as in the metamaterial region. Further inside the metal, similar to the previous examples, we take the limit $\beta_p \to 0$ to model the fact that the wires are severed past the transition layer. In the numerical implementation, the thickness of the metal transition layer was taken equal to $0.04L$.

In Fig. 5 we depict the phase of the reflection coefficient $\rho$ as a function of frequency. Similar to the previous examples (Figs. 2 and 4) it is seen in Fig. 5a that the agreement between the analytical method based on ABCs (solid black curve) and the IDF approach is nearly perfect. The results also concur well with full wave simulations that take into account the granularity of the metamaterial (dashed blue curve). On the other hand, both the DIT1 and the DIT2 approaches yield totally wrong results (Fig. 5b), because they are unable to capture the dynamics of the current along the wires in the vicinity of the ground plane, and the fact that the wires are in ohmic contact with the adjacent region. This not only confirms that a proper discretization is of vital importance at the interfaces, but also shows that the methods DIT1 and DIT2 very inaccurate in a scenario where the double wire medium is attached to a metallic surface.

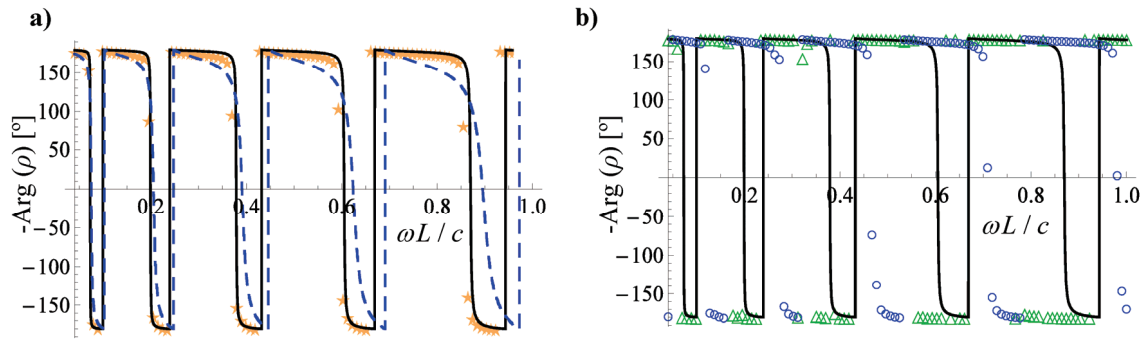

Fig. 5 (Color online) Phase of the reflection coefficient $\rho$ as a function of the normalized frequency for a double wire medium slab with thickness $L$, backed by a PEC surface. Solid (black) curves: mode-matching approach based on additional boundary conditions [11, 12]. (a) Star shaped (orange) symbols:

IDF approach (Sect. III.B); Dashed blue curve: CST Microwave Studio [33]. (b) Triangle shaped (green) symbols: DIT1 approach (Sect. II.B); Circle shaped (blue) symbols: DIT2 approach (Sect. II.B);

We underline that the FDFD implementations do not require any additional boundary conditions, because they assume that Eq. (9) or Eq. (13), depending on the implementation, are valid across the interface. In some sense, as already mentioned in Sec. I, in the FDFD implementations the ABCs are indirectly enforced by the adopted form of the constitutive relation across the interface. For example, the IDF approach, Eq. (13) implicitly imposes that both $P_{c,z}$ and $\frac{1}{\varepsilon_h \beta_p^2}\frac{\partial P_{c,z}}{\partial x}$, with $P_{c,z} = D_z - \varepsilon_0 \varepsilon_h E_z$, are continuous across the interfaces. In case of a wire medium adjacent to a dielectric (e.g. air region) this implies (because we take $\beta_p^2 \to 0$ in the dielectric) that $\partial_x P_{c,z}\big|_{diel} = 0$ at the dielectric side of the boundary. This homogeneous boundary condition will effectively ensure (together with the PML boundary conditions) that $P_{c,z} = 0$ in the dielectric region and thus, because $P_{c,z}$ is continuous at the boundary, that the conduction current vanishes at the wire medium side of the interface, $P_{c,z}\big|_{WM} = 0$, which is equivalent to the ABC used in [12]. On the other hand, if the wire medium is adjacent to a metal transition layer (such that the wires are prolonged into the metal), the continuity of $\frac{1}{\varepsilon_h \beta_p^2}\frac{\partial P_{c,z}}{\partial x}$ enforces that $\partial_x P_{c,z}\big|_{WM} = 0$ at the wire medium side of the boundary, because $\varepsilon_h \to -\infty$ at the metal side. This boundary condition is also equivalent to that considered in Ref. [12]. Thus, it follows that the IDF approach is compatible with the ABC formalism described in our work [12]. It is also interesting to mention that the ABCs implicitly enforced by the DIT1 method are the continuity of $P_{c,z}$ and $\partial_x P_{c,z}$ at the interfaces. These in general are inconsistent with the microstructure of the material because one should have $\partial_x P_{c,z}\big|_{diel} = 0$ and $P_{c,z}\big|_{diel} = 0$ at the dielectric side of the boundary rather than the continuity of $P_{c,z}$ and $\partial_x P_{c,z}$.

The geometries considered in all the previous examples are quite elementary, and due to this reason the considered problems also admit an analytical solution based on mode matching and additional boundary conditions [12]. However, one of the key features of the FDFD-SD approach is that it enables as well to obtain the solution of scattering and waveguiding problems in scenarios wherein electromagnetic waves interact with complex arbitrary shapes of spatially dispersive bodies. Typically, such problems cannot be solved using analytical methods.

To illustrate this, in what follows we investigate the imaging of a source by a metamaterial slab with finite width (inset of Fig. 6a). Previous works [18, 19, 34] have shown that a high-index dielectric material can be used as a lens that enhances the near field and the subwavelength details, and thus enables a superlensing effect. In Refs. [18, 19] it was theoretically suggested and experimentally verified that an ultradense array of crossed metallic wires may have a large index of refraction, and may support highly confined modes with very short propagation wavelengths, which when excited by a source permit restoring the subwavelength spatial spectrum. Next, we study the imaging properties of the double wire medium based on the FDFD-SD (IDF) discretization.

We consider a double wire medium with thickness $L = 10a$ in the near-field of an electric line source placed at a distance $d_1 = 0.04\lambda_0$ above the metamaterial (inset of Fig. 6a). The radius of the wires is $r_w = 0.05a$ and the normalized frequency of operation is $\omega L / c = 0.3$. It is assumed that the wires are PEC and stand in air. The width of the slab along the $y$-direction is $w = 1.2\lambda_0$.

Figure 6a shows the normalized electric field profile at a distance $d_2 = d_1$ below the lens calculated using the FDFD-SD method (star shaped orange symbols), and Fig. 6b shows the associated electric field density plot. The predicted half power beamwidth (HPBW) is $0.13\lambda_0$, which is nearly four times smaller than the traditional diffraction limited value. In the absence of the metamaterial lens, and for the same propagation distance ($d_1+d_2$) in the air region the HPBW would be $0.32\lambda_0$, which clearly confirms that the metamaterial lens can restore the subwavelength details of the source and compensate for the evanescent decay in the air regions. We have also calculated the electric field profile using an analytical model (solid curve in Fig. 6a) based on a Sommerfeld-type integral (see Ref. [18] for details). The analytical method assumes that the metamaterial slab has infinite width *w* along the *y*-direction. As seen in Fig. 6a, the results obtained with the analytical model concur well with the FDFD-SD simulations.

In Figs. 6c and 6d we consider a scenario similar to that that of Figs. 6a and 6b, but in this case the metamaterial lens is illuminated by two electric sources separated by a distance $\Delta_{g,s}=0.25\lambda_0$ (Fig. 6d). It can be seen in Fig. 6c that the metamaterial lens clearly discriminates two sources separated by a distance nearly two times inferior to the diffraction limit. The agreement between the analytical results and the FDFD-SD method is again very satisfactory.

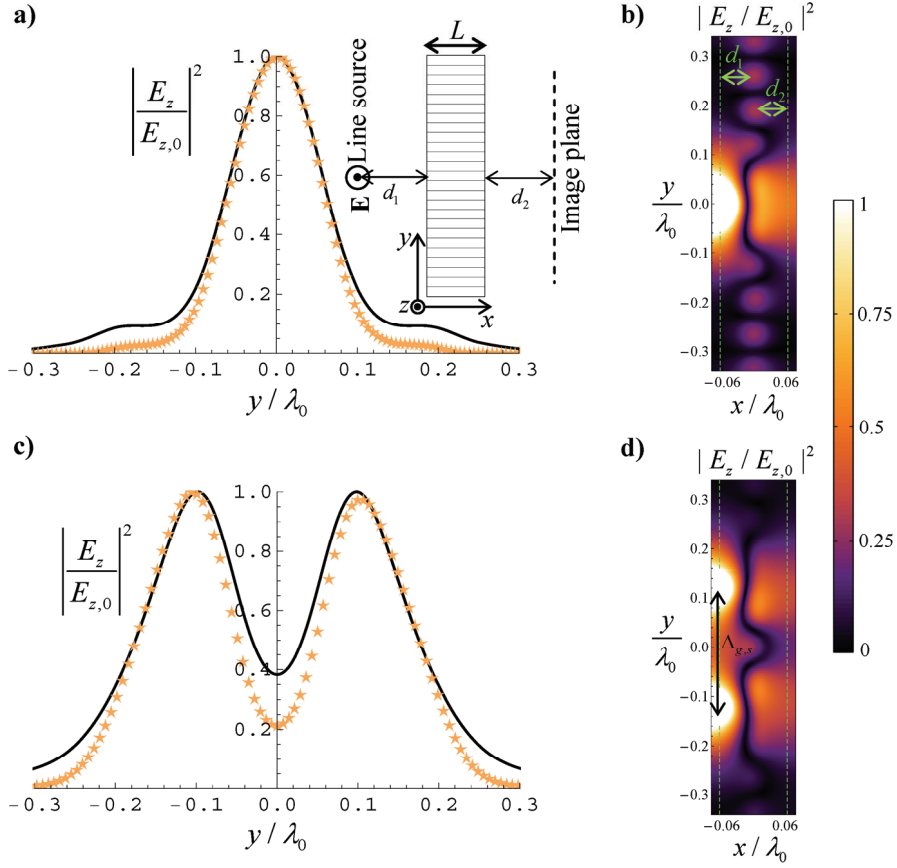

Fig. 6 (Color online) (a) Amplitude of the normalized squared electric field imaged by a metamaterial lens with $L = 0.3\lambda_0 / 2\pi$, $a = L/10$ and $r_w = 0.05a$ (see the inset). Solid curve: analytical model (Ref. [18]). Star shaped orange symbols: FDFD-SD method. (b) Density plot of the normalized electric field for the scenario of panel (a). (c) and (d) similar to (a) and (b), respectively, but for the case wherein the metamaterial lens is illuminated by two electric sources separated by a distance $\Delta_{g,s} = 0.25\lambda_0$.

## V. Concentrating the Electromagnetic Field with a Double Wire Medium Waveguide

It is known that by tapering plasmonic waveguides it may be possible to slow down and ultimately stop the light [35, 36, 37], and concentrate the electromagnetic energy in the nanoscale [35, 38]. In what follows, we show that by tapering a double wire medium waveguide it is possible to enhance significantly the magnetic field toward the tip of the waveguide.

To begin with, we use the FDFD-SD (IDF) method to characterize the guided modes supported by an ultra dense array of metallic wires [21]. Figure 7a shows the dispersion characteristic of the transverse electric (TE) surface wave modes supported by a dense array of PEC wires for different lattice constants *a*. The discrete star-shaped symbols were calculated using the FDFD-SD method and the solid curves were obtained using an analytical method based on mode matching and additional boundary conditions [21]. The dispersion of the guided modes is determined with the FDFD-SD method as follows: For each wavelength of operation ($\lambda_0$) the metamaterial slab is excited by an electric line source placed within the waveguide. Then, the guided wavelength $\lambda_g$ is determined by inspection of the real part of the electric field along the central line of the metamaterial slab (i.e. along the direction of propagation) at a distance sufficiently large (about $0.2\lambda_0$) from the source. The effective index of refraction seen by the guided mode is $n_{eff} = \lambda_0 / \lambda_g = k_y c / \omega$.

Consistent with Ref. [21], Fig. 7a, shows that the metamaterial supports extremely subwavelength guided modes characterized by a large effective index of refraction $n_{eff} = k_y c / \omega$. Moreover, the index of refraction of a guided mode increases as the lattice constant *a* decreases, i.e., as the density of wires increases for a fixed metal volume fraction. The agreement between the results predicted by the numerical method and the analytical model of Ref. [21] is excellent. Figure 7b shows a snapshot in time of the electric field in the *xoy* plane for a double wire medium waveguide with lattice constant $a = L/20$ at the normalized frequency of operation $\omega L / c = 0.1$. As seen, the guided mode is strongly confined to the waveguide, in agreement with the fact that the effective index of refraction is $n_{eff} = 6$ (Fig. 7a).

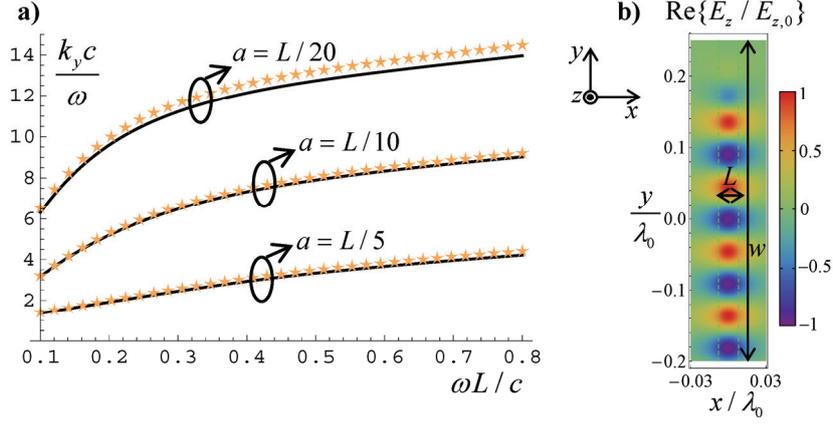

Fig. 7 (Color online) (a) Normalized propagation constant $k_y$ of the TE-guided modes as a function of frequency, for a fixed thickness $L$ of the metamaterial formed by PEC wires, and different lattice constants $a$. The radius of the wires is $r_w = 0.05a$ and the wires stand in air. Solid curve: analytical model (Ref. [21]). Star shaped (orange) symbols: FDFD-SD method. (b) Time snapshot of $E_z$ (in arbitrary unities) at the frequency $\omega L / c = 0.1$ when a waveguide with $a = L/20$ is excited by an electric line source positioned at $(0, -0.2\lambda_0)$.

How can this waveguide be tapered so that the guided electromagnetic energy can be concentrated in an ultra-subwavelength region? To answer this question, first we consider two cascaded waveguides with thickness $L$ and $L_2 = 0.6L$, respectively (Fig. 8a). We want to obtain a matching condition for two waveguides, so that one can ensure a good transmission at the junction. To this end, a transmission line analogy is considered, so that each waveguide is associated with a voltage $V_i$, a current $I_i$ and an impedance $Z_i$ ($i = 1, 2$). To a first approximation, the field component $H_y$ is proportional to the microscopic current flowing in the metallic wires, and thus it should vanish at the interfaces. Thus, from the point of view of the waves *inside* the waveguide, the interfaces with air may be regarded as magnetic walls (PMC). Hence, the guided mode is expected to be quasi-transverse electromagnetic (quasi-TEM) with respect to the direction of propagation (*y*-direction), and that the relevant field components are $E_z$ and $H_x$. Moreover, we can establish the following correspondences:

$$V \sim H_x L, \tag{15a}$$

$$I \sim E_z, \tag{15b}$$

$$Z \sim \frac{H_x}{E_z} L \tag{15c}$$

where $L$ is the thickness of the metamaterial slab along the *x*-direction. Notice that $V$ was associated with $H_x$ and $I$ with $E_z$ because a waveguide with PMC walls is the electromagnetic dual of a standard waveguide with PEC walls. On the other hand, for a TE mode, $H_x \sim \frac{\partial E_z}{\partial y}$ and hence the fields inside the waveguide also satisfy:

$$\frac{H_x}{E_z} \sim k_y. \tag{16}$$

From Eqs. (15c) and (16) it follows that to keep the impedance constant in the two waveguides, and thus ensure a good matching at the transition, one should guarantee that:

$$k_y L = const. \tag{17}$$

Figure 8a shows a density plot of the normalized electric field for a metamaterial waveguide similar to that of Fig. 7a ($a = L/20$) in cascade with another waveguide with thickness $L_2 = 0.6L$. The frequency of operation is $\omega L / c = 0.25$. The lattice constant $a_2$ of the second waveguide is determined so that Eq. (17) is satisfied, i.e, that $k_y L = k_{y,2} L_2$, where $k_y$ and $k_{y,2}$ represent the wave numbers in the waveguide with thickness $L$ and $L_2$, respectively. This can be done by using the analytical model of Ref. [21], provided $k_{y,2}$ and $L_2$ are known. The density plot of Fig. 8a shows that the electric field amplitude is kept nearly constant across the junction of the two waveguides, indicating a good matching. This result is confirmed in Fig. 8b, where the profile of the normalized electric field along the central line of the waveguide is depicted (blue solid curve). It can be seen that despite the abrupt transition, the wave is barely reflected. In contrast, the dashed green curve is obtained without ensuring the impedance match, (specifically the lattice constant $a_2$ is tuned so that $k_y = k_{y,2}$), and in this case a standing wave pattern with a much stronger modulation is obtained.

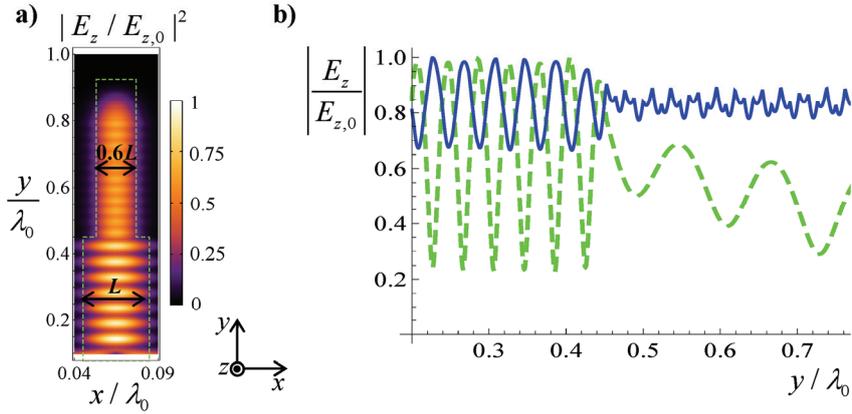

Fig. 8 (Color online) (a) normalized $|E_z|^2$ in the vicinity of two cascaded double wire medium waveguides with thicknesses $L$ and $L_2 = 0.6L$, at the frequency of operation $\omega L / c = 0.25$. The fields were obtained using the FDFD-SD full wave simulator. (b) Profile of the square normalized electric field along the central line of the waveguide. Blue solid curve: the lattice constant in the second waveguide region is tuned so that the impedance matching condition (Eq. 17) is satisfied; Green dashed curve: the lattice constant in the second waveguide region is tuned so $k_y = k_{y,2}$.

Next, we apply this theory to investigate the waveguiding by a tapered metamaterial slab formed by PEC wires with initial thickness $L_i$, that is first tapered toward a tip with thickness $L_f = 0.2L_i$, and then expanded towards its original thickness $L_i$ (see the inset of Fig. 9a). The taper profile is linear and the distance between the points with thickness $L_i$ and $L_f$ is $0.45\lambda_0$. We define $L_{wg} \equiv L_{wg}(y)$ as the thickness of the waveguide as a function of position. The frequency of operation is $\omega L_i / c = 0.25$ and the lattice constant at the beginning of the waveguide is $a_i = L_i / 13$.

Figure 9a shows the effective index of refraction $n_{wg} \equiv k_{y,wg} c/\omega$ seen by the guided mode toward the tip of the waveguide, where $k_{y,wg} \equiv k_{y,wg}(L_{wg})$ is the wave number along the *y*-direction determined so that Eq. (17) is satisfied for each $L_{wg}$. As expected, $n_{wg}$ increases significantly as the tip is approached. In Fig. 9b we depict the lattice constant $a_{wg}$ as a function of the thickness of the waveguide. In the same manner as in Fig. 8, for each $L_{wg}$ the lattice constant $a_{wg}$ is determined so that $k_{y,wg}$ satisfies the matching condition (17). Figure 9c shows a density plot of the normalized electromagnetic fields along the waveguide. Consistent with the results reported in Fig. 8, the electric field remains essentially constant along the waveguide, despite the tapering. This is further supported by Fig. 9d, which shows the normalized electric field profile (blue solid curve) along the axis of the waveguide. The ripple observed in the electric field profile in the vicinity tip may be related to numerical imprecision, as near the tip the guided wavelength is extremely small, and thus a very refined mesh is required to obtain fully converged results. In contrast, both components of the magnetic field are strongly enhanced as the tip is approached, indicating that tapering the metamaterial waveguide permits concentrating the magnetic field into a subwavelength spot (Fig. 9c). This also shown in the inset of Fig. 9d, which depicts $H_x/H_y$ (black curve) and $H_x/H_{x,1}$ (green curve) along the axis of the waveguide, where $H_{x,1}$ is the amplitude of the *x*-component of the magnetic field in a waveguide with constant thickness $L = L_i$. It is evident that $H_y$ is nearly negligible as compared to $H_x$ (black curve), indicating that we have, indeed, quasi-TEM propagation, in agreement with our initial assumption. Moreover, $H_x$ is enhanced about five times with respect to a waveguide with constant thickness $L = L_i$ (green curve), which is consistent with the fact that $L_f/L_i = 5$.

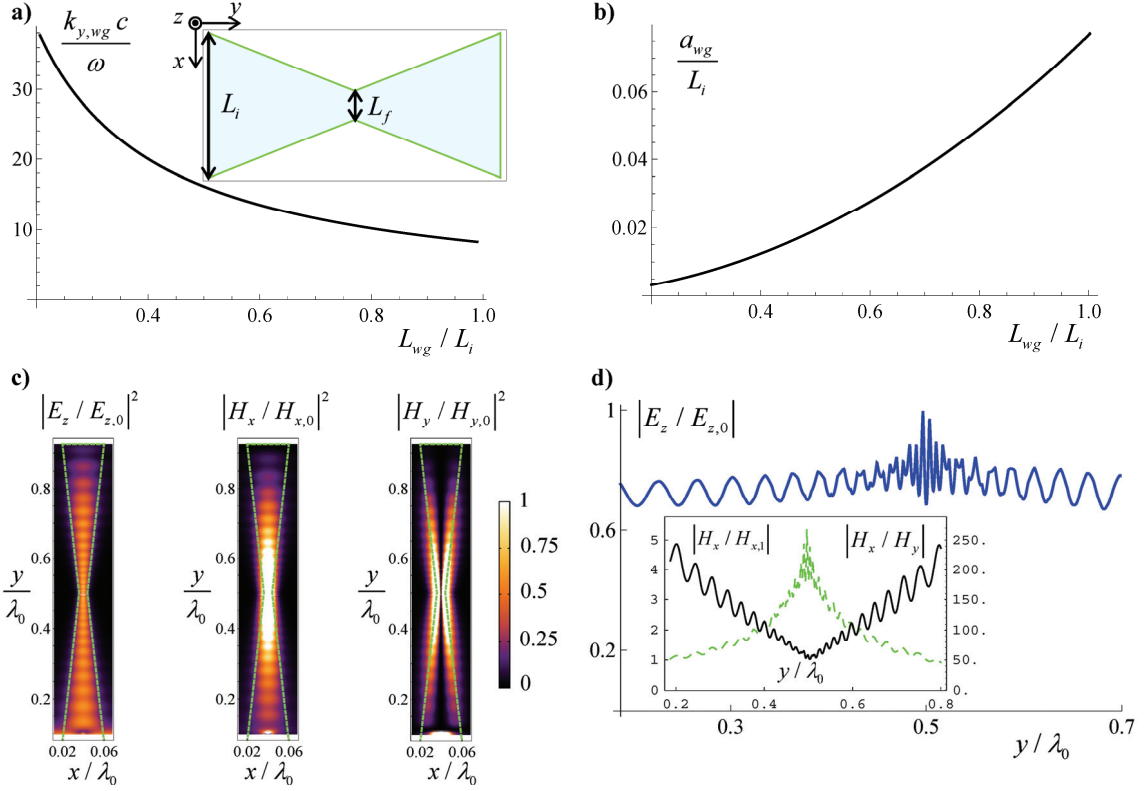

Fig. 9 (Color online) (a) y-component of the guided wave number $k_{y,wg}$ (calculated using Eq. 17) as a function of the normalized thickness of the tapered metamaterial waveguide. The geometry of the waveguide is shown in the inset. (b) Normalized lattice constant $a_{wg}$ as a function of the thickness of the waveguide. (c) Normalized $|E_z|^2$, $|H_x|^2$ and $|H_y|^2$ in the vicinity of the tapered waveguide. (d) Profile of the normalized electric field along the central line of the waveguide. The inset shows the profile of $H_x/H_y$ along the central line of the waveguide (blue solid curve) and $H_x$ normalized to the amplitude of the x-component of the magnetic field in a waveguide with constant thickness $L = L_i$.

We also studied the case where the waveguide is tapered and severed at the tip. In this example the parameters of the waveguide are $a_i = L_i/15$, $L_f = 0.03 L_i$ and the frequency of operation is $\omega L_i / c = 0.15$. Figure 10a shows a density plot of the normalized electromagnetic fields in the vicinity of the tapered waveguide. The results are consistent with the previous example, as the electric field is nearly constant along the waveguide and both components of the magnetic field are greatly enhanced. Figure 10b represents the x-component of the magnetic field along the axis of the waveguide. In agreement with the results of Fig. 9d, the enhancement of the magnetic field is roughly inversely proportional to the compression of the waveguide, confirming that this is an exciting possibility to enhance the magnetic fields in an ultra-subwavelength region.

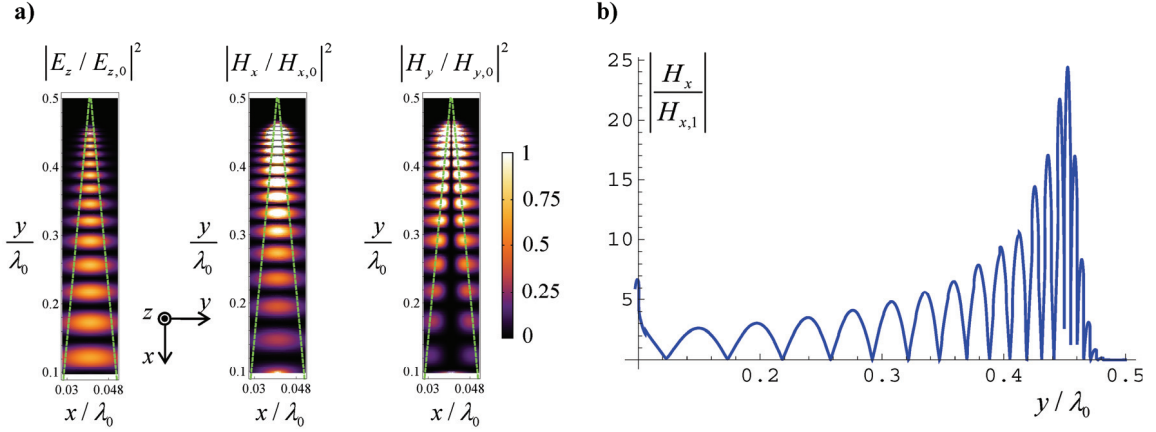

Fig. 10 (Color online) (a) Normalized $|E_z|^2$, $|H_x|^2$ and $|H_y|^2$ in the vicinity of a tapered double wire medium waveguide, with initial and final thicknesses $L_i = 15 a_i$ and $L_f = 0.03 L_i$, respectively. The frequency of operation is $\omega L_i / c = 0.15$. $H_x$ normalized to the amplitude of the x-component of the magnetic field in a waveguide with constant thickness $L = L_i$.

## VI. Conclusions

It was argued that the knowledge of the bulk electromagnetic response of a spatially dispersive material is insufficient to characterize the response to a macroscopic external excitation in presence of interfaces, even in simple scenarios where the geometry of the interfaces is trivial. It was highlighted that the partial differential equations that link **D** and **E**, obtained by inverse Fourier transforming the constitutive relations in the spectral domain, may not hold across a boundary between two different materials, and that it is possible to link **D** and **E** through inequivalent differential equations over the interfaces, but which are totally consistent in the bulk regions. The correct form of the differential equations across the boundary can only be determined based on the knowledge of the internal structure of the metamaterial. It was illustrated how this can be done in practice for the particular case of a double wire medium, and a general FDFD-SD approach was developed to accurately characterize the electromagnetic response of spatially dispersive wire medium bodies with arbitrary geometries. As an application of the developed methods, we investigated the possibility of concentrating the electromagnetic fields at the tip of an ultra compact tapered waveguide formed by wire media, showing that this may be an exciting route for enhancing and focusing the magnetic field in a subwavelength spot.

**Acknowledgments:**

This work was partially supported by Fundação para Ciência e a Tecnologia under project PTDC/EEATEL/100245/2008. J.T.C. acknowledges financial support by IT and Fundação para a Ciência e a Tecnologia under the fellowship SFRH/BD/36976/2007.

# Appendix A

In this Appendix, it is shown that the system (12) can be written exclusively in terms of the electromagnetic fields $E_z$, $D_z$, $H_x$ and $H_y$.

To begin with, we note that from the theory of Refs. [13, 27], the macroscopic response of the wire medium can be described by [this set of equations is represented in a compact manner by Eq. (12)]:

$$\nabla \times \mathbf{E} = i\omega\mu_0 \mathbf{H} \tag{A1}$$

$$\nabla \times \mathbf{H} = \mathbf{j}_{ext} - i\omega\mathbf{D} \tag{A2}$$

$$\frac{\partial}{\partial x_\alpha}\varphi_\alpha = -(Z_w - i\omega L_w)I_\alpha + E_\alpha \tag{A3}$$

$$\frac{\partial}{\partial x_\alpha}I_\alpha = i\omega C_w \varphi_\alpha, \tag{A4}$$

$$\mathbf{D} = \varepsilon_0 \varepsilon_h \mathbf{E} + \frac{1}{-i\omega} \sum_\alpha \frac{I_\alpha}{A_{cell}} \hat{\mathbf{u}}_\alpha \tag{A5}$$

where $A_{cell} = a^2$, $\hat{\mathbf{u}}_\alpha$ is a unit vector that defines the orientation of the $\alpha$-th set of wires ($\alpha=1,2$), $C_w$, $L_w$ and $Z_w$ are the effective capacitance, inductance and self-impedance of the wires per unit length of a wire, respectively [13, 27], $\mathbf{j}_{ext}$ represents an external excitation, and the second term in the right-hand side of Eq. (A5) is the macroscopic density of current associated with flow of charges along the metallic wires, $\mathbf{J}_w = \sum_\alpha \frac{I_\alpha}{A_{cell}} \hat{\mathbf{u}}_\alpha$. In the above, $I_\alpha$ and $\varphi_\alpha$ are the current and additional potential associated with the $\alpha$-th set of wires, and $E_\alpha = \hat{\mathbf{u}}_\alpha \cdot \mathbf{E}$ $x_\alpha = \hat{\mathbf{u}}_\alpha \cdot \mathbf{r}$ with $\mathbf{r} = (x,y,z)$. Thus, substituting Eq. (A3) into Eq. (A4) one finds that:

$$C_w \frac{\partial}{\partial x_\alpha} \left( \frac{1}{C_w} \frac{\partial I_\alpha}{\partial x_\alpha} \right) + \left( \frac{\omega}{c} \right)^2 \varepsilon_h I_\alpha + i\omega C_w Z_w I_\alpha = i\omega C_w E_\alpha \tag{A6}$$

where we used the fact that $C_w L_w = \varepsilon_0 \varepsilon_h \mu_0$ for the case of straight wires [27]. For the configuration of interest in this work, we know that both the electric displacement vector and the electric field only have a *z*-component. Therefore, for propagation in the *xoy* plane we may write:

$$E_\alpha = \mathbf{E} \cdot \hat{\mathbf{u}}_\alpha = E_z \hat{\mathbf{z}} \cdot \hat{\mathbf{u}}_\alpha = E_z \frac{1}{\sqrt{2}} \qquad (\alpha=1,2) \tag{A7}$$

$$\frac{I_\alpha}{A_{cell}} = \mathbf{J}_w \cdot \hat{\mathbf{u}}_\alpha = J_{w,z} \hat{\mathbf{z}} \cdot \hat{\mathbf{u}}_\alpha = J_{w,z} \frac{1}{\sqrt{2}} \qquad (\alpha=1,2), \tag{A8}$$

and by substituting Eqs. (A7) and (A8) into Eq. (A6) we obtain

$$C_w \frac{\partial}{\partial x_\alpha} \left( \frac{1}{C_w} \frac{\partial J_{w,z}}{\partial x_\alpha} \right) + \left( \frac{\omega}{c} \right)^2 \varepsilon_h J_{w,z} + i\omega C_w Z_w J_{w,z} = \frac{1}{A_{cell}} i\omega C_w E_z . \tag{A9}$$

On the other hand, $\frac{\partial}{\partial x_\alpha} = \hat{\mathbf{u}}_\alpha \cdot \nabla$ and since we assume $\frac{\partial}{\partial z} = 0$, this implies that

$\frac{\partial}{\partial x_\alpha} = \pm \frac{1}{\sqrt{2}} \frac{\partial}{\partial x}$. Hence, we finally obtain the result:

$$\frac{1}{2} C_w \frac{\partial}{\partial x}\left(\frac{1}{C_w}\frac{\partial J_{w,z}}{\partial x}\right) + \left(\frac{\omega}{c}\right)^2 \varepsilon_h J_{w,z} + i\omega C_w Z_w J_{w,z} = \frac{1}{A_{cell}} i\omega C_w E_z \qquad (A10)$$

Using now Eq. (A5) and the definition of $\mathbf{J}_w$, it follows that $D_z = \varepsilon_0 \varepsilon_h E_z - \frac{J_{w,z}}{i\omega}$ and hence:

$$\frac{C_w}{2}\frac{\partial}{\partial x}\left[\frac{1}{C_w}\frac{\partial}{\partial x}\left(\varepsilon_h E_z - \frac{D_z}{\varepsilon_0}\right)\right] + \varepsilon_h\left(\frac{\omega}{c}\right)^2\left(\varepsilon_h E_z - \frac{D_z}{\varepsilon_0}\right) + i\omega C_w Z_w \left(\varepsilon_h E_z - \frac{D_z}{\varepsilon_0}\right) = \frac{C_w}{A_{cell}\varepsilon_0} E_z$$
(A11)

Finally, we use $\beta_p^2 = \frac{C_w}{\varepsilon_0 \varepsilon_h A_{cell}}$ and $Z_w = -\frac{1}{i\omega \pi r_w^2 \varepsilon_0 \varepsilon_h (\varepsilon_m/\varepsilon_h - 1)} = \frac{\beta_c^2}{i\omega C_w}$ where

$\beta_c^2 = -\frac{\beta_p^2}{(\varepsilon_m/\varepsilon_h - 1) f_V}$ [13, 27] to rewrite Eq. (A11) as in Eq. (13a) of the main text. On

the other hand, (13b) follows directly from Eqs. (A1)-(A2).

# Appendix B

Here we provide explicit formulas for the discretized system (13a)-(13b):

$$\frac{\varepsilon_h(i,j)\beta_p^2(i,j)}{2}\left[\frac{\varepsilon_h^{-1}(i,j)\beta_p^{-2}(i,j) + \varepsilon_h^{-1}(i+1,j)\beta_p^{-2}(i+1,j)}{2\Delta x^2} P_{c,z}(i+1,j) - \right.$$
$$\frac{2\varepsilon_h^{-1}(i,j)\beta_p^{-2}(i,j) + \varepsilon_h^{-1}(i+1,j)\beta_p^{-2}(i+1,j) + \varepsilon_h^{-1}(i-1,j)\beta_p^{-2}(i-1,j)}{2\Delta x^2} P_{c,z}(i,j) +$$
$$\left.\frac{\varepsilon_h^{-1}(i,j)\beta_p^{-2}(i,j) + \varepsilon_h^{-1}(i-1,j)\beta_p^{-2}(i-1,j)}{2\Delta x^2} P_{c,z}(i-1,j)\right] \qquad (B1)$$
$$+\left[\varepsilon_h(i,j)\left(\frac{\omega}{c}\right)^2 + \beta_c^2(i,j)\right] P_{c,z}(i,j) + \varepsilon_0\varepsilon_h(i,j)\beta_p^2(i,j) E_z(i,j) = 0$$

$$\frac{E_z(i+1,j)-2E_z(i,j)+E_z(i-1,j)}{\Delta x^2}+\frac{E_z(i,j+1)-2E_z(i,j)+E_z(i,j-1)}{\Delta y^2}+ \left(\frac{\omega}{c}\right)^2 \frac{D_z(i,j)}{\varepsilon_0}=-i\omega\mu_0 j_{s,z}(i,j) \quad (B2)$$